\documentclass[11pt,a4paper]{article}
\usepackage[margin=25mm]{geometry}
\usepackage[T1]{fontenc}
\usepackage[utf8]{inputenc}
\usepackage{lmodern}
\usepackage[nopatch=footnote]{microtype}
\usepackage{amsmath,amssymb,amsthm}
\usepackage{graphicx,subcaption,booktabs,tabularx,longtable,array}
\usepackage{dirtytalk}
\usepackage{authblk}
\usepackage{placeins}
\usepackage[hidelinks]{hyperref}
\usepackage{xurl}
\usepackage{setspace}
\newcommand{\AOnePercent}{17.6}
\newcommand{\AParity}{50.3}
\newcommand{\AFinalOne}{0.096}
\newcommand{\AFinalTwo}{0.890}
\newcommand{\BFinalMinor}{0.0335}
\newcommand{\BFinalIncumbent}{0.274}
\newcommand{\CPopOne}{0.461}
\newcommand{\CPopTwo}{0.456}
\newcommand{\CEquilibrium}{0.4575}
\newcommand{\CMaxEigen}{-0.0010}
\newcommand{\BridgeMAE}{0.00163}
\newcommand{\BridgeMAESD}{0.00107}
\newcommand{\BridgeMAEMin}{0.00036}
\newcommand{\BridgeMAEMax}{0.00434}
\newcommand{\BridgeR}{0.99879}
\newcommand{\BridgeRSD}{0.00181}
\newcommand{\BridgeRMin}{0.99155}
\newcommand{\BridgeRMax}{0.99996}
\newcommand{\HorizonDifference}{37.3}
\newcommand{\SweRecruitmentR}{0.813}
\newcommand{\SweRecruitmentMAE}{5.275}

\newcommand{\SwePiecewiseR}{0.996}
\newcommand{\SwePiecewiseMAE}{0.809}
\newcommand{\SwePiecewiseHoldout}{2.244}

\newcommand{\SweLinearHoldout}{12.771}
\newcommand{\TurLogisticR}{0.918}
\newcommand{\TurLogisticMAE}{0.312}
\newcommand{\TurLogisticHoldout}{0.723}

\newcommand{\TurLinearHoldout}{0.325}
\newcommand{\NZRecruitmentR}{0.968}
\newcommand{\NZRecruitmentMAE}{1.166}
\newcommand{\NZRecruitmentHoldout}{1.197}

\newcommand{\NZLinearHoldout}{0.302}

\title{An Epidemiological Modeling Take on Religion Dynamics}
\author[1]{Bilge Taskin\thanks{These authors contributed equally.}}
\author[1,2]{Teddy Lazebnik\textsuperscript{*}\thanks{Corresponding author: \texttt{teddy.lazebnik@ju.se}.}}
\affil[1]{Department of Computing, J\"onk\"oping University, J\"onk\"oping, Sweden}
\affil[2]{Department of Information Systems, University of Haifa, Haifa, Israel}
\date{}
\begin{document}
\maketitle
\begin{abstract}
Religions are among the most consequential social institutions, shaping collective identities, moral norms, and political organization across societies and historical periods. Nevertheless, despite extensive scholarship describing conversion, competition, and secularization, there is still no widely adopted formal model that captures religious dynamics over time within a unified, mechanistic framework. In this study, we propose an epidemiologically grounded model of religious change in which religions spread and compete analogously to co-circulating strains. The model extends multi-strain compartmental dynamics by distinguishing passive believers, active missionaries, and religious elites, and by incorporating demographic turnover and mutation-like splitting that endogenously generates new denominations. Using computer simulations, we show that the same mechanism reproduces canonical qualitative regimes, including emergence from rarity, rapid expansion, long-run coexistence, and transient rise-and-fall movements. A reduced calibration variant fits historical affiliation trajectories with parsimonious regime shifts in effective recruitment and disaffiliation, yielding interpretable signatures of changing social conditions. Finally, sensitivity analyses map sharp regime boundaries in parameter space, indicating that modest shifts in recruitment efficacy or retention among active spreaders can qualitatively alter long-run religious landscapes. These results establish a general, interpretable framework for studying religion as a dynamical diffusion process and provide a tool for comparative inference and counterfactual analysis in sociological research.
\end{abstract}

\section{Introduction}
\label{sec:introduction}
Religion has played a central role in shaping human societies, influencing cultural identity, political institutions, and collective behavior throughout history \cite{intro_1,intro_2,intro_3}. In early agricultural societies, animistic and polytheistic traditions provided explanatory frameworks for natural phenomena and mechanisms for social cohesion \cite{intro_4,intro_5,intro_6}. The institutionalization of priesthoods and ritual practices in Mesopotamia, Egypt, and the Indus Valley can be seen as early examples of religion serving as a stabilizing force for centralized states, legitimizing political authority and codifying moral order \cite{intro_7,intro_8,intro_9}. Later, the emergence of axial-age religions such as Judaism, Buddhism, Confucianism, and early Greek philosophy introduced universal moral codes and abstract cosmologies, enabling broader social integration and the consolidation of empires \cite{intro_10,intro_11,intro_12}. Following this in the historical record, in the classical and medieval periods, the spread of Christianity and Islam illustrates how religious movements can transcend ethnic and geographic boundaries, often accompanying processes of imperial expansion, trade, and conquest \cite{intro_13,intro_14,intro_15}. Religious institutions became central to governance, education, and law, shaping not only the political order but also economic practices such as taxation, charity, and land distribution \cite{intro_16,intro_17}. Nowadays, modernity has not diminished religion's influence, but transformed its modalities of spread and competition. The Enlightenment and secularization movements challenged traditional institutions, yet religion adapted by giving rise to new denominations, evangelical movements, and, in some cases, politically mobilized fundamentalisms \cite{intro_18,intro_19,intro_20}. In the globalized era, religions circulate alongside cultural products and ideas, facilitated by digital media and migration \cite{intro_21,intro_22}. Contemporary dynamics include the expansion and diversification of Pentecostal and charismatic movements across national boundaries \cite{intro_23,intro_24,intro_25}. At the same time, disaffiliation and nonreligious identification have emerged as significant counter-trends in many industrialized societies \cite{intro_26}.

These historical patterns strongly suggest that religions function analogously to evolving systems of ideas that compete for adherents and adapt to shifting socio-political environments \cite{intro_27}. Crucially, religion not only spreads within societies but also reshapes them, influencing demographic patterns, institutional stability, and economic organization \cite{intro_28}. For this reason, the spread of religion has long been a subject of scholarly attention across disciplines \cite{intro_29,intro_30,intro_31,intro_32,Jacquet2021Religiosity}. Historians and sociologists have examined large-scale processes of conversion, conflict, and reform; anthropologists have investigated the ritual, symbolic, and cultural mechanisms that facilitate belief transmission; and political scientists have studied the role of religion in shaping governance, legitimacy, and social movements. 

Alongside this scholarship, formal models have examined church growth, new religious movements, evolutionary diversification, and changes in religiosity \cite{intro_33,upal2005simulating,religion_sir,gore2018forecasting}. These studies motivate examining how competition, behavioral roles, and schism interact within one transparent dynamical structure. To address this limitation, recent research has drawn attention to the structural similarities between the spread of religions and the spread of infectious diseases \cite{intro_33,intro_35,Carrignon2019CulturalTransmission}. Both involve populations of susceptible individuals, mechanisms of contact that facilitate transmission, processes of \say{infection} (conversion), and \say{recovery} (disaffiliation or apostasy). Furthermore, just as pathogens may coexist, compete, or mutate, so too can religions diversify, fragment, and evolve.

This analogy naturally motivates borrowing compartmental mechanisms from epidemiology, including the Susceptible--Infected--Recovered framework \cite{first_sir}, without assuming that religious affiliation follows literal disease states. In this work, we introduce a multi-religion model distinguishing passive believers, active missionaries, and a protected institutional role. Conversion, disaffiliation, competition, and schism-like relabeling can generate different trajectories under explicitly specified settings. We also define an optional link between external covariates and conversion rates, while keeping that untested extension separate from the experiments actually performed.

This paper makes three primary contributions. First, we present a role-structured diffusion model with a fixed set of competing affiliations. Second, we provide reproducible deterministic and stochastic experiments, mechanism ablations, and a local coexistence analysis that distinguish endogenous schism from externally forced transients. Third, we assess what reduced historical curve fits can and cannot establish, using documented observation definitions and temporal holdouts. These fits illustrate aggregate descriptions; they do not calibrate or validate the full multi-religion framework or identify causal social and political effects.

The remainder of this paper is organized as follows. Section~\ref{sec:rw} reviews cultural and religious diffusion and relevant epidemiological models. Section~\ref{sec:model} defines the states, transitions, and optional contextual-rate function. Section~\ref{sec:experiments} describes the simulations, historical observations, and sensitivity design. Section~\ref{sec:results} reports the resulting trajectories, empirical fitting and holdout performance, and numerical agreement. Section~\ref{sec:dicussion} discusses interpretation, limitations, and future research; the Appendix provides computational and calibration details.

\section{Related Work}
\label{sec:rw}
In this section, we review the underlying religion spread dynamics that one should consider in any modeling effort, followed by a short review of epidemiological modeling practices with a focus on extended SIR models with multi-strain. 

\subsection{Religion spread dynamics}
Religion has long been recognized as one of the most enduring and influential forces in human societies \cite{cretacci2003religion,ross1896social,kaasa2013religion}. It provides not only systems of belief but also shared practices, moral codes, and institutional structures that bind communities together and regulate social behavior \cite{rw_1}. Scholars across sociology, anthropology, history, and political science have examined religion as both a driver and a product of social change, noting its capacity to shape collective identities, stabilize political orders, and motivate large-scale cooperation \cite{rw_2,rw_3,rw_4,rw_5}. At the same time, religion has been a source of competition, division, and conflict, often catalyzing transformative cultural and political shifts \cite{rw_6,rw_7}.

The spread of religion is not simply a matter of individual belief but a complex socio-cultural process influenced by institutions, social networks, and political environments \cite{Tverskoi2023Beliefs}. Early anthropological theories emphasized religion as a form of \say{social glue}, binding communities through ritual and shared narratives \cite{rw_8}. In agricultural and early state societies, religious practices often reinforced stratified political hierarchies: priesthoods legitimized rulers, while rulers endowed temples, intertwining spiritual authority with political power \cite{rw_9}. The integration of religion into law and governance provided durable mechanisms for transmitting norms across generations and stabilizing expanding polities \cite{haggard2010governance}.

At a macro-historical scale, religion often spread through contact processes that resemble contagion: through trade, conquest, migration, and missionary activity \cite{intro_21}. The expansion of Buddhism along the Silk Road, Christianity throughout the Roman Empire, and Islam across North Africa and into Asia are notable examples where the growth of religious communities followed routes of human mobility and exchange \cite{rw_10,rw_11}. In such contexts, social prestige, institutional support, and political patronage frequently amplified the rate of conversion, while persecution and legal restrictions could slow or suppress diffusion. Thus, the likelihood of religious adoption was never purely a matter of theology but deeply conditioned by social and political context \cite{intro_29}.

Another key feature of religious spread is competition among belief systems. Rarely do religions exist in isolation; instead, they compete for adherents within overlapping populations. This competition can manifest as coexistence (as in multi-religious empires), dominance by a single tradition (as in state religions), or dynamic cycles of conversion and reform. Religious competition has historically led to both violent conflict (such as the Crusades or sectarian wars) and peaceful coexistence under frameworks of tolerance or pluralism \cite{finke2016consequences,fox2020world,rw_14,rw_15}. Notably, this competitive aspect mirrors ecological and epidemiological systems, where multiple strains or species vie for limited resources.

Religions also split and diversify, creating new sects, denominations, or entirely new movements \cite{rw_12,rw_13}. Schisms within Christianity, the formation of Sunni and Shia traditions in Islam, or the branching of Hindu and Buddhist schools illustrate how internal doctrinal disputes, leadership crises, or external pressures can fragment a religion into competing sub-groups. These splits often lead to rapid bursts of growth for the new branch, particularly when aligned with favorable social or political conditions. Such processes resemble mutation in biological systems, generating diversity within a lineage that can alter the competitive balance among strains.

Growth, institutionalization, fragmentation, and decline are additional patterns in religious history \cite{rw_12,rw_13}. New religious movements often begin as small, tightly bound communities, sometimes dismissed as cults or sects, but may grow rapidly when their message resonates with marginalized or disaffected groups \cite{rw_16}. Over time, some of these movements institutionalize and become dominant traditions, while others decline, fragment, or disappear altogether. The rise-and-fall cycles of millenarian cults, revivalist movements, and reformist sects highlight the temporal dynamics of religious spread and decline, often linked to broader socio-political turbulence.

Taken together, these insights point toward a few essential features of religion spread that are particularly relevant for formal modeling. First, religions diffuse through contact-like processes, influenced by networks of social interaction and mobility. Second, they compete as multiple strains within the same population, where outcomes can include coexistence, dominance, or extinction. Third, religions are subject to mutation-like splitting, producing new branches that alter the competitive landscape. Fourth, their dynamics can exhibit rise-and-fall patterns, reflecting interactions between belief systems and broader socio-political forces. Finally, the rate and direction of spread are strongly modulated by external conditions---such as political stability, institutional regulation, and cultural diversity---which motivate explicit contextual extensions and appropriate empirical tests.

\subsection{Epidemiological modeling}
There are many types of modeling methods suggested in the literature to capture pandemic spread dynamics \cite{tang2020review,chang2020game,pare2020modeling}. The simplest type of models are functional (phenomenological) models, which include logistic, Richards, Gompertz, and other growth curves fitting \cite{Lazebnik2023EcoModReview}. Other models focus on a bottom-to-top approach, simulating individual-level interactions to capture population-level spread dynamics with methods like the agent-based simulation \cite{Lazebnik2024ABMsupply,kerr2021covasim,silva2020covid,hunter2020hybrid}. Data-driven methods also include physics-informed neural networks that combine differential-equation structure with observed trajectories \cite{PLOSCompBio2025VOCsINN}. Nevertheless, compartment models, such as the SIR model, remain widely used because their compartmental mechanisms are explicit and can be confronted with observations \cite{zhang2022usage,mirsaeedi2025compartmental}. In these models, one partitions the population into epidemiological states and describes flows between them via differential (or difference) equations \cite{first_sir}. 

A foundational example is the classical SIR model, which decomposes the population into three compartments: $S(t)$, $I(t)$, and $R(t)$ with dynamics described by the following system of ordinary differential equations (ODEs): 
\begin{equation}
\label{eq:sir}
\begin{aligned}
\frac{dS}{dt} = -\beta \frac{S I}{N}, \; \; \;
\frac{dI}{dt} = \beta \frac{S I}{N} - \gamma I,  \; \; \;
\frac{dR}{dt} = \gamma I,
\end{aligned}
\end{equation}
where $N=S+I+R$ is constant, $\beta>0$ is the effective transmission rate, and $\gamma>0$ is the recovery/removal rate \cite{first_sir}. This model is deliberately simplified and does not capture all real-world epidemiological processes. As such, many extensions have been proposed over the years that extend it to capture more realistic epidemiological dynamics. Most extensions are obtained by relaxing its simplifying assumptions, so it is useful to view them systematically. Common extensions: (i) add infection stages (e.g., exposed/latent, presymptomatic, asymptomatic) to represent progression and infection-to-infectiousness delays \cite{Hethcote2000}; (ii) distinguish outcomes of infection (e.g., hospitalization, severe disease, death) as separate compartments rather than absorbing them into a single removed class \cite{Hethcote2000}; (iii) introduce additional time scales via demographic turnover, waning immunity, or time-varying transmission such as seasonality \cite{Hethcote2000}; and (iv) relax homogeneous mixing by incorporating population structure (age, risk, space), yielding coupled compartmental systems \cite{shuchami2025spatio}.

These mechanisms can first be developed for a single pathogen and then extended to competing strains. Real-world pandemics frequently involve multiple co-circulating strains (or variants) whose epidemiological parameters and immune-escape properties differ over time \cite{shami2022economic}. Extending single-strain compartmental models to multi-strain settings can be done at several levels of resolution, ranging from minimal \say{parallel-strain} models
to history-based constructions that track immune states induced by prior infections.

A generic strategy is to represent strain identity explicitly (e.g., by duplicating infected compartments) and to
encode immunological interactions using either (i) complete cross-immunity, (ii) partial cross-immunity through
a susceptibility reduction matrix, or (iii) explicit immune-history compartments indexed by the set (or sequence)
of previously experienced strains \cite{lazebnik2022generic}.

The most parsimonious multi-strain extension assumes that infection by any strain removes an individual from
future susceptibility (i.e., no reinfection by any strain). Let $I_i(t)$ denote the number infectious with strain
$i\in\{1,\dots,m\}$, with strain-specific transmission and recovery rates $(\beta_i,\gamma_i)$. A natural model is
\begin{equation}
\label{eq:ms_sir_no_reinf}
\begin{aligned}
\frac{dS}{dt} &= -\frac{S}{N}\sum_{i=1}^m \beta_i I_i,\\
\frac{dI_i}{dt} &= \beta_i \frac{S I_i}{N} - \gamma_i I_i, \qquad i=1,\dots,m,\\
\frac{dR}{dt} &= \sum_{i=1}^m \gamma_i I_i,
\end{aligned}
\end{equation}
where $N=S+\sum_i I_i+R$ is constant. This formulation captures competition for a shared susceptible pool and
represents competition during a transient outbreak, but does not by itself establish persistent endemic coexistence. Models that relax complete cross-protection can instead track partial cross-immunity and infection history \cite{Andreasen1997Cocirculating,GogGrenfell2002ManyStrain,MinayevFerguson2009Realism}.

Allowing reinfection requires relaxing the assumption that recovery permanently removes individuals from the
susceptible class. Two common mechanisms are waning immunity and incomplete cross-immunity. In the first, individuals who recovered return to susceptibility over time while in the latter, prior infection reduces but does not eliminate susceptibility to the same or other strains. A convenient reduced description introduces strain-specific recovered classes $R_i$ and permits infection of
previously recovered individuals at a reduced rate. For example, letting $\sigma_{ij}\in[0,1]$ denote the relative
susceptibility to strain $j$ after recovery from strain $i$, a schematic reinfection term takes the form
$\beta_j \sigma_{ij} \, R_i I_j/N$. In its most general form, reinfection leads to immune-history indexing
(e.g., $R_J$ for individuals recovered from the set of strains $J$), which scales combinatorially with the number
of strains and naturally encodes infection order \cite{lazebnik2022generic}.
Such models are useful when repeated infections materially affect population-level dynamics, including thresholds
in which reinfection becomes dominant \cite{GomesWhiteMedley2005ReinfectionThreshold}, strain invasion and
coexistence regimes \cite{NunesTeloDaGamaGomes2006LocalizedContacts}, and settings where a newly emerging
strain can effectively infect individuals with prior immunity \cite{FudoligHoward2020LocalStability}.

Multi-strain models can treat strain appearance as exogenous where new strains are introduced at specified times or
endogenous where new strains are generated by mutation from existing strains. A common endogenous representation is a mutation flow between infectious classes:
\begin{equation}
\label{eq:mutation_flow}
\frac{dI_i}{dt} = \cdots - \sum_{j\neq i}\mu_{ij} I_i + \sum_{j\neq i}\mu_{ji} I_j,
\end{equation}
where $\mu_{ij}$ is the rate at which infections of strain $i$ generate infections of strain $j$ (e.g., via within-host
mutation and onward transmission). For many-strain settings, mutation is often coupled to antigenic distance so that both (a) mutation transitions and (b) cross-immunity depend on a strain-space metric. This yields a bridge between discrete multi-strain compartment models and evolutionary/antigenic-drift descriptions
\cite{AbuRaddadFerguson2004CrossImmunityMutation,GordoGomesReisCampos2009GeneticDiversity,MeehanRojekEvers2018CoupledMultiStrain}.

\section{Model Definition}
\label{sec:model}
\begin{figure}[!htbp]
\centering\includegraphics[width=.94\linewidth]{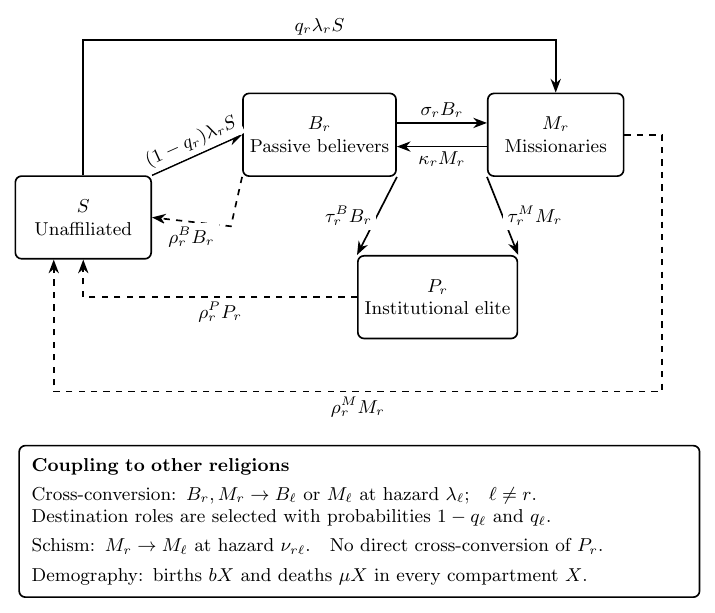}
\caption{Role-structured transitions for a focal religion $r$. Conversion is generated by missionaries; a new convert enters $B_r$ with probability $1-q_r$ or $M_r$ with probability $q_r$. Cross-conversion and missionary relabeling couple this structure to the other religions, as specified below the diagram. Births and deaths apply to every compartment.}
\label{fig:model_scheme}
\end{figure}

We model religious affiliation as a multi-strain, contact-driven diffusion process in which religions compete for adherents. The epidemiological analogy specifies transition mechanisms; it does not identify religious belief with pathology. The model combines role heterogeneity, cross-affiliation conversion, demographic turnover, and missionary relabeling representing schism within a fixed set of potential religions \cite{lazebnik2022generic,shami2022economic}. Unlike classical SIR, disaffiliation returns individuals to a susceptible state, and there is no permanently recovered compartment. Figure~\ref{fig:model_scheme} summarizes the transitions.

Formally, let $\mathcal R=\{1,\ldots,m\}$ be a fixed set of modeled religions; some may initially be absent. Each person occupies exactly one state: $S$ (unaffiliated), $B_r$ (affiliated but not actively proselytizing), $M_r$ (actively proselytizing), or $P_r$ (an institutional or religious elite protected from direct cross-conversion). Thus
\begin{equation}
N(t)=S(t)+\sum_{r=1}^m[B_r(t)+M_r(t)+P_r(t)].
\end{equation}
The model excludes simultaneous affiliations. The distinction between $B_r$, $M_r$, and $P_r$ is behavioral and institutional, not a measured distinction in the historical datasets used below.

Under homogeneous mixing, the force of conversion to religion $r$ is
\begin{equation}
\lambda_r(t)=\beta_r(t)\frac{M_r(t)}{N(t)},\qquad
X_r(t)=S(t)+\sum_{j\ne r}[B_j(t)+M_j(t)].
\label{eq:force}
\end{equation}
Here $\beta_r$ is an effective contact-and-conversion rate per missionary, not a probability conditional on a measured contact. All continuous-time rates have units year$^{-1}$; $q_r\in[0,1]$ is dimensionless. Conversion into religion $r$ occurs at rate $\lambda_rX_r$ and is divided between $B_r$ and $M_r$ with weights $1-q_r$ and $q_r$. Each $B_r$ and $M_r$ individual faces an outward cross-conversion hazard $\sum_{\ell\ne r}\lambda_\ell$. Priests do not directly generate conversions and cannot be directly converted to another religion, although they may disaffiliate.

Within an affiliation, $\sigma_r$ governs $B_r\to M_r$, $\kappa_r$ governs $M_r\to B_r$, and $\tau_r^B,\tau_r^M$ govern progression into $P_r$. Disaffiliation from role $X\in\{B,M,P\}$ occurs at rate $\rho_r^X$. Missionary relabeling from $r$ to $\ell$ occurs at rate $\nu_{r\ell}\geq0$, with $\nu_{rr}=0$. This transfers an existing individual from $M_r$ to $M_\ell$; it does not create a person or an additional equation. It represents entry into a previously specified denomination, rather than unrestricted innovation of new labels.

Births and deaths occur at rates $bX$ and $\mu X$ in every compartment $X$. This composition-preserving assumption is a demographic bookkeeping approximation: it does not assert that newborns literally become missionaries or priests. It implies $\dot N=(b-\mu)N$. Uniform demographic growth cancels from the deterministic equations for population fractions, although demographic sampling contributes finite-population variation in the stochastic implementation.

The resulting ODE system is
\begin{align}
\dot S&=(b-\mu)S-S\sum_k\lambda_k+
 \sum_k(\rho_k^BB_k+\rho_k^MM_k+\rho_k^PP_k),\label{eq:S-dynamics}\\
\dot B_r&=(b-\mu)B_r+(1-q_r)\lambda_r X_r-B_r\sum_{\ell\ne r}\lambda_\ell
 -\sigma_rB_r+\kappa_rM_r-(\rho_r^B+\tau_r^B)B_r,\label{eq:B-dynamics}\\
\dot M_r&=(b-\mu)M_r+q_r\lambda_rX_r-M_r\sum_{\ell\ne r}\lambda_\ell
 +\sigma_rB_r-(\kappa_r+\rho_r^M+\tau_r^M)M_r\nonumber\\
&\quad+\sum_{\ell\ne r}\nu_{\ell r}M_\ell-\sum_{\ell\ne r}\nu_{r\ell}M_r,\label{eq:M-dynamics}\\
\dot P_r&=(b-\mu)P_r+\tau_r^BB_r+\tau_r^MM_r-\rho_r^PP_r.\label{eq:P-dynamics}
\end{align}
All transfers are nonnegative at the boundary of the state space, so nonnegative initial states remain nonnegative.

\paragraph{Optional contextual modulation.}
To make the proposed connection to external conditions explicit, a positive-rate extension is
\begin{equation}
\beta_r(t)=\beta_{r,0}\exp\{\boldsymbol\eta_r^\mathsf{T}\mathbf z_r(t)\},
\label{eq:context}
\end{equation}
where $\mathbf z_r(t)$ contains specified social, institutional, or political covariates and $\boldsymbol\eta_r$ their coefficients. The implementation accepts supplied covariate series, with linear interpolation between supplied times. This extension is motivated by context-sensitive social-epidemiological modeling \cite{emit_paper}; it is not estimated or empirically tested here. The experiments instead use constant rates, or explicitly prescribed piecewise-constant rates in Scenario B and the reduced Sweden fit. No measured political effect, or causal interpretation of a fitted rate change, is claimed.

\paragraph{Role of the historical reduction.}
For a single religion, setting $B=P=0$, $q=1$, all role-transition rates and mutation to zero, and $y=M/N$ gives exactly
\begin{equation}
\dot y=\beta(t)y(1-y)-\rho(t)y.
\label{eq:histfit_reduced_ode}
\end{equation}
These restrictive assumptions remove the mechanisms that distinguish the full model. More generally, $\dot y$ depends on the unobserved missionary share and role-specific losses, so aggregate affiliation alone does not close the full system. We use Eq.~\eqref{eq:histfit_reduced_ode} as a descriptive reduced benchmark, not as empirical validation of role heterogeneity, competition, or schism. For publisher activity and no-religion reporting, its coefficients are phenomenological rather than literal missionary conversion and apostasy rates.
\section{Model Analysis}
In this section, we analyze the proposed model through a sequence of experiments designed to evaluate both its
descriptive realism and its explanatory and predictive utility. We first outline the experimental design, including the simulation protocol, parameterization strategy, and the summary statistics used to compare model outputs
to empirical patterns. We then present results from three complementary experiment types: (i) stylized forward
simulations demonstrating that the model reproduces qualitative religion dynamics commonly observed in real populations (e.g., emergence, rapid expansion, coexistence, replacement, and persistence under competition); (ii)
historical case studies showing that the model can be calibrated to match documented shifts in religious composition and that the inferred mechanisms provide interpretable explanations for the observed transitions; and (iii) a
sensitivity analysis mapping how key parameters govern regime changes, enabling controlled movement between
distinct dynamical outcomes.

\subsection{Experimental design}
\label{sec:experiments}
We distinguish three tasks: controlled simulations of the full model, descriptive historical fits of reduced curves, and a finite-horizon sensitivity map. All simulated rates and horizons are expressed in years. The stylized parameterizations are illustrative, not calibrated historical time scales. We use $N_0=10{,}000$, $b=\mu=0.01$, and stochastic step $\Delta t=0.1$ year. Each main scenario has 100 independent replicates. The deterministic system is solved with adaptive DOP853, relative and absolute tolerances $10^{-9}$, with integration restarted from the evolved endpoint at every scheduled event. Details of the exchangeable-agent stochastic implementation, seeds, software versions, and numerical checks appear in the Appendix.

\subsubsection{Stylized religious dynamics}
\begin{table}[!htbp]\centering\small
\caption{Executable behavioral parameters. All entries except $q$ have units year$^{-1}$. B's last two columns give each minor group's rates before and after its own scheduled change. Unlisted demographic settings are $b=\mu=0.01$ in all scenarios.}\label{tab:scenario_params}
\setlength{\tabcolsep}{6pt}\begin{tabular}{lcccccc}\toprule
Parameter & A: $r=1$ & A: $r=2$ & B: $r=1$ & B: minor, pre & B: minor, post & C: both\\\midrule
$\beta$ & 0.03 & 0.45 & 0.04 & 0.75 & 0.02 & 0.28\\
$q$ & 0.05 & 0.1 & 0.05 & 0.02 & 0.02 & 0.05\\
$\sigma$ & 0.02 & 0.03 & 0.02 & 0.005 & 0.005 & 0.018\\
$\kappa$ & 0.01 & 0.01 & 0.02 & 0.02 & 0.02 & 0.018\\
$\tau^B$ & 0.002 & 0.003 & 0.002 & 0.0003 & 0.0003 & 0.0015\\
$\tau^M$ & 0.003 & 0.004 & 0.003 & 0.0003 & 0.0003 & 0.002\\
$\rho^B$ & 0.005 & 0.004 & 0.006 & 0.01 & 0.05 & 0.006\\
$\rho^M$ & 0.004 & 0.003 & 0.006 & 0.01 & 0.15 & 0.006\\
$\rho^P$ & 0.001 & 0.001 & 0.001 & 0.01 & 0.4 & 0.001\\
\bottomrule\end{tabular}\end{table}

\begin{table}[!htbp]\centering\small
\caption{Initial counts and coupling. Counts for B are specified immediately before its year-zero introduction. All scenarios start with $N_0=10{,}000$, and all omitted mutation entries are zero.}\label{tab:initials}
\begin{tabular}{lccccc}\toprule
Scenario & $S_0$ & $(B_1,M_1,P_1)$ & $(B_2,M_2,P_2)$ & $T$ (years) & Mutation (year$^{-1}$)\\\midrule
A & 6000 & $(3000,200,800)$ & $(0,0,0)$ & 216 & $\nu_{12}=0.002$\\
B & 5000 & $(4200,300,500)$ & $(0,0,0)$ & 216 & None\\
C & 6300 & $(2000,120,180)$ & $(1200,80,120)$ & 1000 & $\nu_{12}=\nu_{21}=0.02$\\
\bottomrule\end{tabular}\end{table}

\textbf{Scenario A (schism-mediated emergence and replacement).}
Religion 1 is the only initially present religion, with 40\% of the population affiliated; 60\% is unaffiliated. Religion 2 starts absent and is seeded solely by missionary relabeling at $\nu_{12}=0.002$ year$^{-1}$. Its conversion rate is $0.45$, compared with $0.03$ for religion 1. All other settings are listed in Tables~\ref{tab:scenario_params}--\ref{tab:initials}. We simulate 216 years and compare the result with a no-schism ablation that changes only $\nu_{12}$ to zero.

\textbf{Scenario B (externally introduced, forced transient movements).}
Five initially absent minor religions are introduced sequentially by transfers from $S$, without mutation. Each begins with 50 passive believers and 100 missionaries. At its separately prescribed change time, its conversion rate decreases from $0.75$ to $0.02$, while its disaffiliation rates increase (Table~\ref{tab:scenario_params}). Table~\ref{tab:schedules} specifies every event. This experiment asks whether the same compartmental structure represents growth followed by decline under explicit external forcing; it does not test spontaneous emergence or endogenous recurrent waves. Ablations remove either all introductions or all subsequent rate changes.
\begin{table}[!htbp]\centering\small
\caption{Scenario B's externally prescribed events. Each introduction transfers 50 people from $S$ to $B_r$ and 100 from $S$ to $M_r$, preserving $N$. At the change year, only the four rates indicated in Table~\ref{tab:scenario_params} change.}\label{tab:schedules}
\begin{tabular}{ccc}\toprule Minor religion & Introduction year & Rate-change year\\\midrule
2 & 0 & 40\\
3 & 16 & 60\\
4 & 48 & 90\\
5 & 112 & 145\\
6 & 167 & 190\\
\bottomrule\end{tabular}\end{table}

\textbf{Scenario C (symmetric, mutation-supported coexistence).}
The religions have identical behavioral parameters but unequal initial prevalence. Symmetric missionary relabeling is $\nu_{12}=\nu_{21}=0.02$ year$^{-1}$. We integrate to 1000 years to examine relaxation, not to make a millennium-scale social forecast. We locate a coexistence equilibrium and evaluate the Jacobian after eliminating $S$ using the population-sum constraint. Negative real parts of all six remaining eigenvalues establish local, not global, asymptotic stability. A separate ablation reduces both relabeling rates to $0.0002$; symmetry of behavioral parameters is not assumed sufficient for balanced coexistence.

\subsubsection{Historical records and observation definitions}
\label{subsec:hist_fit}
\textbf{Sweden (membership decline).}
We use the Church of Sweden's official membership percentages for 1972--2024, excluding 1981, for which the source has no population-register figure: 52 observations in total \cite{svenskakyrkan_membership_1972_2024}. The observed variable is membership divided by population, not the missionary compartment. A constant-rate reduced model is compared with a four-rate specification using a fixed breakpoint at 2000. The breakpoint is a prespecified descriptive comparison; its timing does not identify the effect of any particular institutional change. The source also records changes in membership rules, underscoring that these rates aggregate more than interpersonal conversion.

\textbf{Turkey (active publisher reporting).}
We use annual peak Jehovah's Witness publisher counts divided by the population reported in the corresponding service-year table, 2014--2024 (11 observations) \cite{jw_service_year_report_2024,jw_country_reports_2024,jw_historical_reports,upstream_code}. Peak and annual-average publisher counts are not mixed. These organizational reports measure active publishing rather than all adherents, so they are not observations of $B+M+P$. We fit the two-parameter logistic description $\dot y=r y(1-y/K)$, where $r$ is an effective growth rate and $K$ is an asymptotic share. We do not separately fit three redundant parameters in $\beta y(K-y)-\rho y$, since that equation identifies at most $\beta K-\rho$ and $\beta$ from an ideal trajectory. We compare the logistic curve with anchored linear and exponential trends, without an additional fitted breakpoint.

\textbf{New Zealand (no-religion reporting).}
We assemble no-religion counts for the 2001, 2006, 2013, 2018, and 2023 censuses from Stats NZ releases \cite{statsnz2019religion,statsnz2024religion}. We divide each by the corresponding \emph{all-resident census count}, not a sum of potentially overlapping affiliation categories or a mixture of response-based denominators \cite{teara2001population,statsnz2013population,statsnz2018population,statsnz2024religion}. Table~\ref{tab:nzdata} supplies all five numerators and denominators. Consequently, the complement includes unclassified or nonresponding residents as well as people reporting a religion. This is a descriptive two-category reporting series, not a closed observation of religious versus nonreligious switching. Common denominator definitions do not eliminate changes in response, imputation, or census processing. We also fit the separately retained published percentage series as a sensitivity check, without mixing it into the primary count ratio. No breakpoint is fitted to five observations.

\paragraph{Fitting and temporal holdouts.}
All curves start at the first observed value. Dynamical parameters are fitted by bounded nonlinear least squares, using 32 deterministic multi-start initializations for the nonlinear two- and four-parameter fits. Exact continuous segment solutions replace numerical integration for these reduced equations. The Sweden state is propagated continuously through 2000 and is never reset to an observed breakpoint value. Because the first point is fixed, all reported fitting metrics exclude it, leaving 51, 10, and 4 informative residuals respectively. An anchored linear trend has one fitted slope; the Turkish exponential comparison has one fitted rate. All comparisons minimize squared errors on the original share scale. Bounds, diagnostic AICc values, and the exact solution are given in the Appendix.

We also fit only observations through 2014 for Sweden, 2021 for Turkey, and 2018 for New Zealand, and evaluate the remaining 10, 3, and 1 observations without updating parameters or reinitializing states. These are illustrative historical holdouts, not prospective preregistered forecasts. Their small number, especially one NZ point, limits generalization. No uncertainty intervals for historical mechanism parameters are asserted; boundary solutions and sensitivity to the reporting definition are reported rather than interpreted as precise causal estimates.

\subsubsection{Finite-horizon sensitivity}
\label{subsec:heatmap}
We vary Scenario A's $\beta_2$ over 21 equally spaced values in $[0.03,0.60]$ and $\rho_2^M$ over 21 values in $[0.001,0.15]$, keeping all other settings fixed. Each of the 441 configurations is solved deterministically. We report the continuous share $y_2(216)$ and retain $y_2(432)$ to assess horizon dependence. There are no threshold-assigned ``takeover'' or ``coexistence'' labels, and the map does not estimate takeover probabilities or establish a bifurcation.

\subsection{Results}
\label{sec:results}
\begin{figure}[!htbp]\centering
\begin{subfigure}[t]{.48\textwidth}\includegraphics[width=\linewidth]{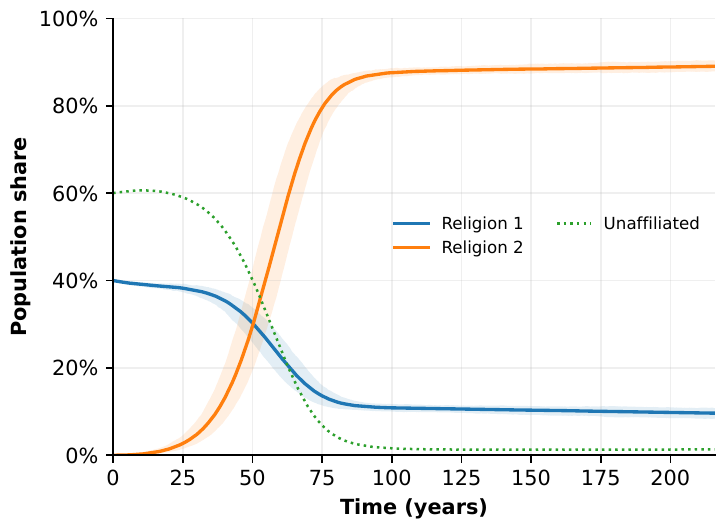}\caption{Schism-mediated replacement.}\end{subfigure}\hfill
\begin{subfigure}[t]{.48\textwidth}\includegraphics[width=\linewidth]{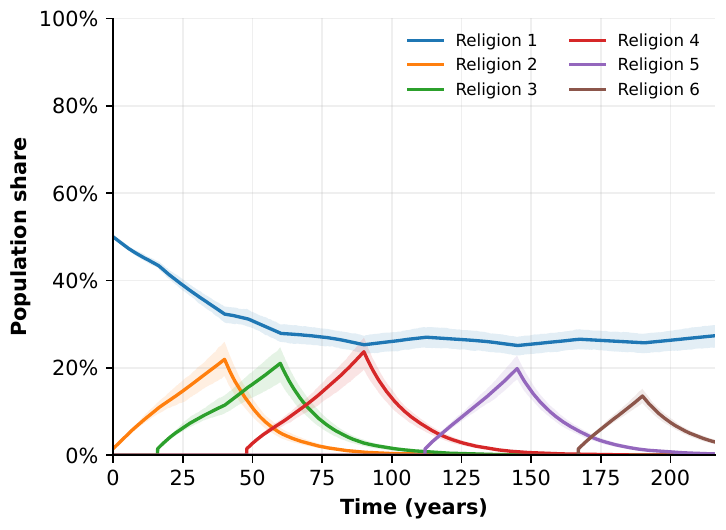}\caption{Externally forced transients.}\end{subfigure}\par\medskip
\begin{subfigure}[t]{.48\textwidth}\includegraphics[width=\linewidth]{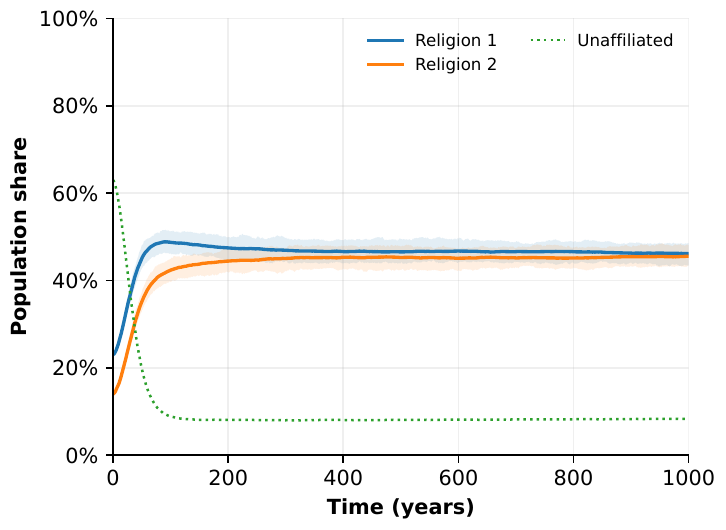}\caption{Mutation-supported coexistence.}\end{subfigure}
\caption{Full-model stochastic simulations (100 replicates per scenario). Lines are replicate-mean shares and shaded bands are pointwise central 95\% replicate intervals, not confidence intervals for historical predictions. Scenario B uses the complete forcing schedule in Table~\ref{tab:schedules}. Time is in years, with the longer horizon in C chosen for relaxation.}
\label{fig:new_religion_threepanel}\end{figure}

Figure~\ref{fig:new_religion_threepanel} shows mean trajectories with pointwise 2.5th--97.5th percentiles across stochastic runs. In Scenario A, religion 2 first exceeds 1\% at year \AOnePercent{} and reaches parity with religion 1 at year \AParity{}. At year 216, mean shares are $y_1=\AFinalOne$ and $y_2=\AFinalTwo$. Removing mutation leaves religion 2 identically absent. Thus its establishment in this experiment is genuinely schism-mediated, while its subsequent expansion follows from the assigned conversion and retention advantage.

In Scenario B, each minor religion peaks at its externally specified rate-change time. Peak mean shares range from approximately 0.136 to 0.237, and the combined minor share at year 216 is \BFinalMinor{}. The incumbent ends at \BFinalIncumbent{}, rather than remaining constant. Removing introductions leaves every minor religion absent; removing the rate changes instead allows the earliest minor religion to persist at a terminal deterministic share of approximately 0.651. The short-lived waves depend on the prescribed schedule. They are neither evidence of autonomous oscillations nor a prediction of movement-specific collapse dates.

In Scenario C, terminal mean shares are \CPopOne{} and \CPopTwo{}. The symmetric deterministic equilibrium has $y_1=y_2=\CEquilibrium$ and all six constrained Jacobian eigenvalues are negative, with largest real part \CMaxEigen{} year$^{-1}$. This supports locally stable coexistence for the specified parameterization. With the much weaker symmetric mutation rate $0.0002$, the unequal initialization instead yields deterministic terminal shares of approximately 0.905 and 0.031 at year 1000. Balanced coexistence is a property of the specified coupling and role structure, not a consequence of parameter symmetry alone. The equilibrium also allocates a large fraction to the protected $P$ role; these uncalibrated role proportions should not be read as realistic priesthood prevalence.

\begin{figure}[!htbp]\centering
\begin{subfigure}[t]{.48\textwidth}\includegraphics[width=\linewidth]{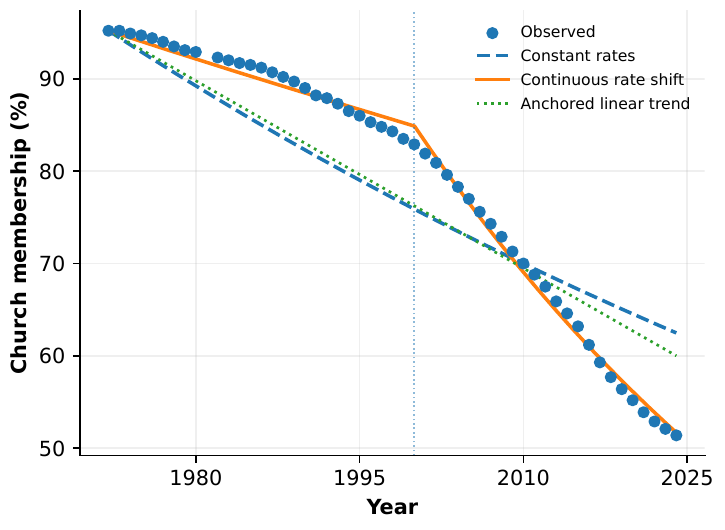}\caption{Sweden: membership.}\end{subfigure}\hfill
\begin{subfigure}[t]{.48\textwidth}\includegraphics[width=\linewidth]{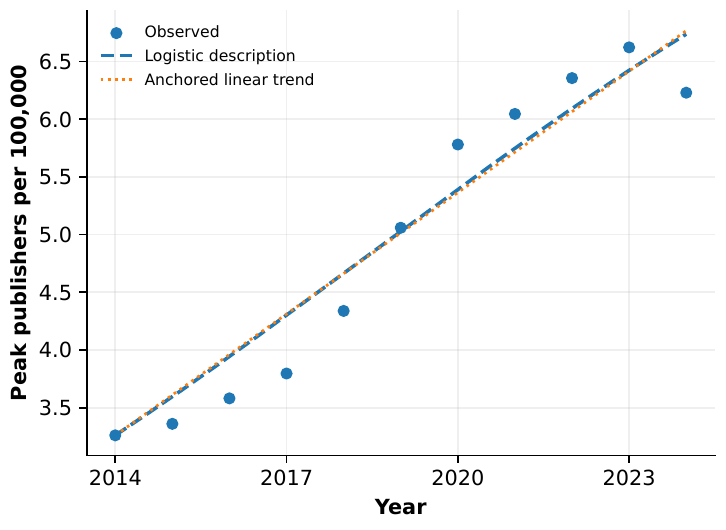}\caption{Turkey: peak publishers.}\end{subfigure}\par\medskip
\begin{subfigure}[t]{.48\textwidth}\includegraphics[width=\linewidth]{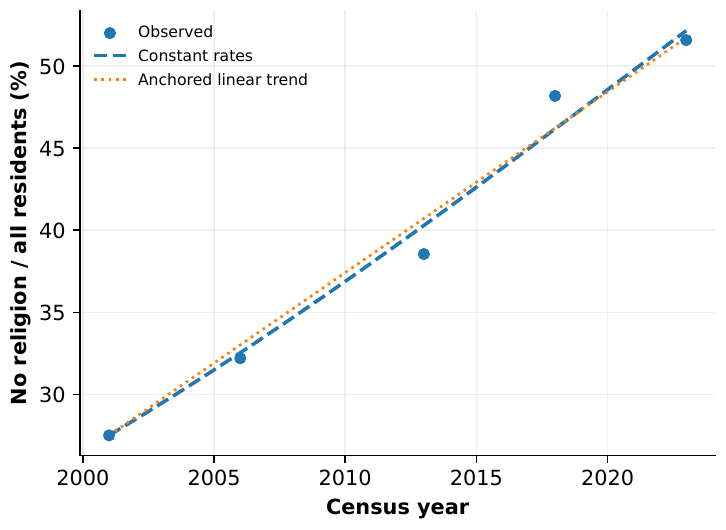}\caption{NZ: no-religion reporting.}\end{subfigure}
\caption{Observed records and full-sample descriptive fits. The observables differ across cases and are not treated as interchangeable measurements of the full model. NZ uses no-religion counts divided by all census residents. Dashed/dotted comparisons and legends identify the fitted specifications; the Sweden change is continuous at 2000. Held-out forecasts are shown separately in the Appendix.}
\label{fig:historic_fit}\end{figure}

Figure~\ref{fig:historic_fit} and Table~\ref{tab:fitmetrics} separate fitted agreement from held-out errors. For Sweden, the fixed-break, continuous rate-shift fit improves $R^2$ from \SweRecruitmentR{} to \SwePiecewiseR{}, and MAE from \SweRecruitmentMAE{} to \SwePiecewiseMAE{} percentage points. The fitted recruitment rates lie at the lower bound in both segments; the effective loss rate increases from approximately 0.00408 to 0.02068 year$^{-1}$. This is a descriptive summary of declining membership, not evidence that recruitment literally ceased. Training through 2014 gives a held-out MAE of \SwePiecewiseHoldout{} percentage points for the rate-shift model, compared with \SweLinearHoldout{} for the anchored linear trend.

For Turkey, the logistic description attains $R^2=\TurLogisticR$ and MAE \TurLogisticMAE{} peak publishers per 100,000 residents. It summarizes the broad increase but cannot reproduce the observed 2024 decline as an actual downturn. Its held-out MAE is \TurLogisticHoldout{} publishers per 100,000, compared with \TurLinearHoldout{} for the linear trend. Thus the additional saturation parameter does not establish a predictive advantage on this holdout.

New Zealand's constructed no-religion share increases at every census; it exhibits no rise-and-fall episode. The constant-rate fit has $R^2=\NZRecruitmentR$ and MAE \NZRecruitmentMAE{} percentage points. Its 2023 holdout error is \NZRecruitmentHoldout{} percentage points, versus \NZLinearHoldout{} for the linear trend. Fitting the separately published percentage series instead raises in-sample $R^2$ to approximately 0.993 and changes the estimated rates, illustrating sensitivity to response denominators and data processing. Neither fit validates conversion mechanisms, and near-perfect interpolation of five points is not sought. Table~\ref{tab:historical_params} reports every dynamical parameter, including both Sweden periods.

\begin{table}[!htbp]\centering\small
\caption{Historical fitting and held-out errors. Sweden and NZ MAE values are percentage points; Turkey MAE values are publishers per 100,000 residents. $n$ excludes the fixed initial observation and $k$ counts fitted curve parameters. Holdouts are 2015--2024 (Sweden), 2022--2024 (Turkey), and 2023 (NZ). Lower MAE is better.}\label{tab:fitmetrics}
\begin{tabular}{llccccc}\toprule Case & Specification & $n$ & $k$ & In-sample $R^2$ & Fit MAE & Holdout MAE\\\midrule
Sweden & Constant rates & 51 & 2 & 0.813 & 5.275 & 14.327\\
Sweden & Rate shift & 51 & 4 & 0.996 & 0.809 & 2.244\\
Sweden & Linear trend & 51 & 1 & 0.861 & 4.590 & 12.771\\
Turkey & Logistic & 10 & 2 & 0.918 & 0.312 & 0.723\\
Turkey & Exponential & 10 & 1 & 0.887 & 0.336 & 0.723\\
Turkey & Linear trend & 10 & 1 & 0.910 & 0.330 & 0.325\\
NZ & Constant rates & 4 & 2 & 0.968 & 1.166 & 1.197\\
NZ & Linear trend & 4 & 1 & 0.961 & 1.272 & 0.302\\
\bottomrule\end{tabular}\end{table}

\begin{table}[!htbp]\centering\small
\caption{All fitted parameters of the historical dynamical curves. Values below $10^{-8}$ are shown as zero (boundary solutions). Rate parameters are year$^{-1}$; $K$ and $y_0$ are population fractions. The Sweden rate shift is fixed at 2000 and the simulated state remains continuous there.}\label{tab:historical_params}
\begin{tabular}{llccccc}\toprule Case & Specification/period & $\beta$ & $\rho$ & $r$ & $K$ & $y_0$\\\midrule
Sweden & Constant & 0 & 0.0080957 & -- & -- & 0.952\\
Sweden & Pre-2000 & 0 & 0.0040809 & -- & -- & 0.952\\
Sweden & Post-2000 & 0 & 0.020682 & -- & -- & 0.952\\
Turkey & Logistic & -- & -- & 0.15241 & 9.5816e-05 & 3.2606e-05\\
NZ & Constant & 0.047966 & 0 & -- & -- & 0.27508\\
\bottomrule\end{tabular}\end{table}

\begin{figure}[!htbp]\centering\includegraphics[width=.76\linewidth]{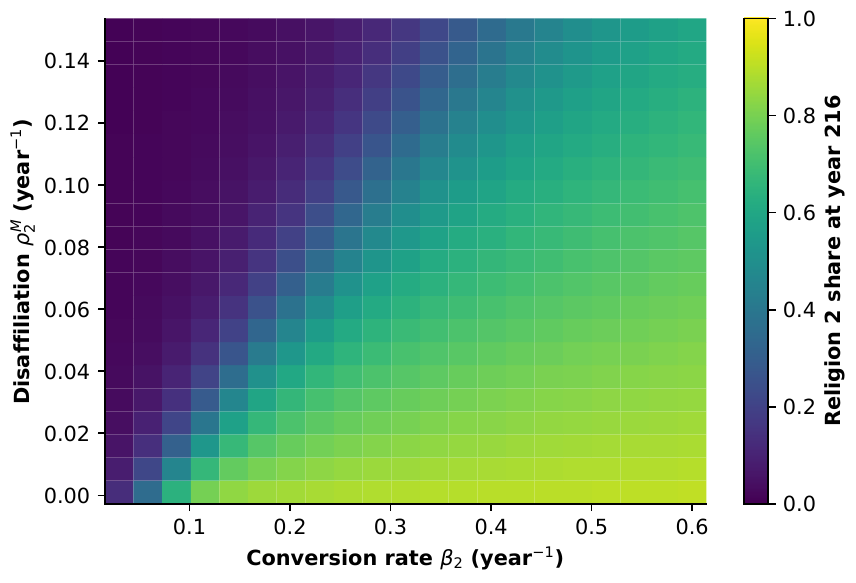}
\caption{Deterministic finite-horizon sensitivity of Scenario A on a $21\times21$ grid. Color denotes $y_2(216)$, not a categorical regime or takeover probability. All rates are annual, and other parameters are fixed. The companion data include year-432 values to expose horizon sensitivity.}
\label{fig:splitting_regimes}\end{figure}

Figure~\ref{fig:splitting_regimes} displays continuous finite-horizon changes in religion 2's prevalence as conversion and missionary disaffiliation vary. Greater recruitment and lower loss generally favor expansion in this configuration, but a visible gradient is not itself a phase transition. Doubling the horizon changes the terminal share by as much as \HorizonDifference{} percentage points across the grid. It would be misleading to assign universal long-run regimes using a single finite-horizon threshold.

\begin{figure}[!htbp]\centering\includegraphics[width=.76\linewidth]{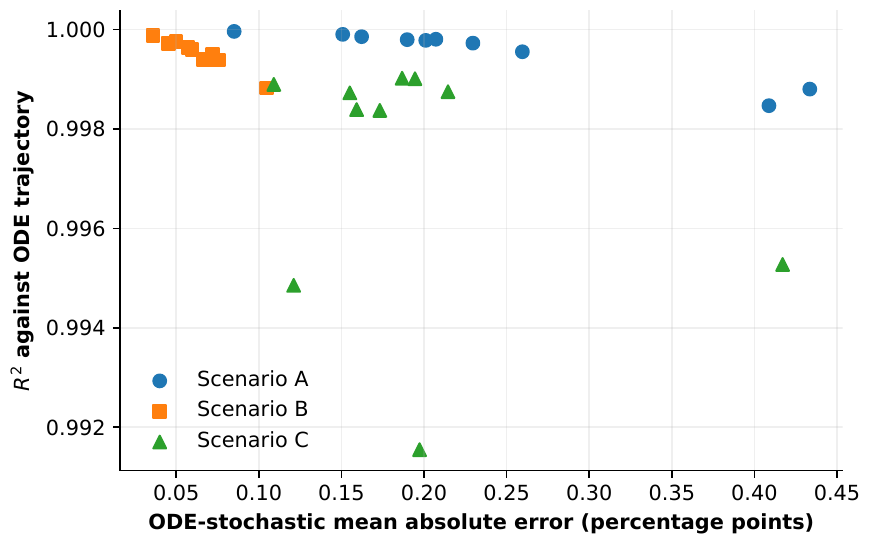}
\caption{ODE versus stochastic replicate-mean agreement for 30 configurations, with 20 replicates each. Each marker averages MAE and $R^2$ over that configuration's religions; marker shape identifies the source scenario. MAE on the horizontal axis is expressed in percentage points. Mutation is included in A and C, not B.}
\label{fig:bridge_full}\end{figure}

Across 30 perturbed configurations (ten per scenario), using 20 stochastic replicates per configuration, mean religion-averaged MAE is $\BridgeMAE\pm\BridgeMAESD$ and mean $R^2$ is $\BridgeR\pm\BridgeRSD$; the uncertainty here is the sample standard deviation across configurations. The respective ranges are [\BridgeMAEMin, \BridgeMAEMax] and [\BridgeRMin, \BridgeRMax]. Figure~\ref{fig:bridge_full} displays exactly these 30 case-level observations. A and C include mutation; B has none and retains its scheduled introductions and rate changes. These checks concern implementation agreement, not empirical validity.

\section{Discussion}
\label{sec:dicussion}
In this study, we introduce an epidemiologically inspired framework for modeling religious dynamics as a multi-strain diffusion process with role heterogeneity and schism-like missionary relabeling. Building on classical compartment modeling and cultural-diffusion perspectives, the model formalizes contact-driven conversion, institutional retention, and cross-affiliation competition. We evaluate its internally generated dynamics separately from externally forced experiments and from reduced descriptive fits to historical records. An explicit contextual-rate extension is specified, but no social or political coefficient is estimated in the present study.

Our results show that the proposed structure can represent emergence and replacement, externally induced rise-and-fall trajectories, and mutation-supported coexistence under different settings. These outcomes are consistent with the range of adoption trajectories discussed in diffusion and cultural transmission research \cite{rogers2003diffusion,cavalliSforzaFeldman1981cultural,boydRicherson1985culture}, without establishing that the selected parameters explain a particular historical mechanism. The ablations sharpen this distinction: Scenario A requires incoming schism to seed an absent challenger; Scenario B requires imposed introductions and subsequent adverse rate changes to produce the displayed waves; and Scenario C's balanced coexistence depends on its coupling strength. In contrast to threshold-based complex-contagion models \cite{centolaMacy2007complex,reisinger2024patterns}, the present homogeneous-mixing conversion hazard is linear in missionary prevalence and contains no social-reinforcement threshold. The finite-horizon sensitivity map demonstrates parameter sensitivity, not complex contagion or a proven phase transition \cite{fullsack2021regime}.

Interpreted sociologically, the parameters provide a compact language for expressing competition in a shared adherent pool and organizational differences in recruitment and retention \cite{iannaccone1991market}. The historical fits show that Sweden's membership decline can be summarized more accurately by a continuous change in effective rates than by a stationary specification. However, the aggregate data do not distinguish interpersonal conversion from cohort replacement, administrative membership changes, or migration. The Turkish publisher and NZ reporting series further demonstrate why descriptive fit should not be equated with forecasting: simpler linear trends achieve lower held-out errors in those two cases. These observations are compatible with, but do not independently test, wider accounts of institutional change and existential security \cite{berger1967sacredcanopy,norrisInglehart2012sacred,gore2018forecasting}.

The framework may support sociological inquiry in three ways. First, competing hypotheses can be mapped to explicit parameter changes and evaluated using observable predictions and ablations, rather than curve resemblance alone. Second, comparative analysis could relate estimated rates to education, state policy, media exposure, or social heterogeneity when harmonized data and identifiability permit \cite{rogers2003diffusion,norrisInglehart2012sacred}. Such an analysis would require measuring the covariates in Eq.~\eqref{eq:context}, specifying an observation model, and testing against appropriate simpler alternatives. Third, the full model provides a conditional simulator for exploring changes in conversion opportunities, retention, or schism propensity. These simulations are mathematical scenarios, not identified historical counterfactuals or policy-effect estimates. This distinction fits the broader use of epidemic-style formalisms for rumors and information diffusion \cite{daleyKendall1965rumours,kabir2019analysis,zhao2013sir,rui2018spir,matsumoto2025information}.

This study has several limitations. First, homogeneous mixing omits structured networks, spatial heterogeneity, and reinforcement; network or spatial extensions could represent these processes more directly \cite{centolaMacy2007complex,zheng2023spatiotemporal,mahmood2022modeling}. Mobility and migration can also interact with religious change, but are not included here \cite{intro_21,ng2022muslim}. Second, exclusive affiliations omit syncretism and dual practice, and the fixed label set limits the meaning of endogenous innovation. Third, the illustrative settings were selected to explore particular behaviors, not estimated from role-resolved observations. The protected-role accumulation in Scenario C, copying of behavioral roles by the demographic bookkeeping scheme, and the imposed events in B preclude literal demographic interpretation. Fourth, historical fits calibrate reduced curves only; they neither identify the full model nor compare its extra compartments empirically with simpler mechanisms. Low-frequency NZ data, denominator and census-processing differences, organization-reported publisher counts, boundary optima, and small temporal holdouts further restrict inference. Richer observables, cohort and migration data, and an explicit measurement-error model are required before uncertainty in mechanism parameters or cross-context prediction can be assessed reliably \cite{cavalliSforzaFeldman1981cultural,norrisInglehart2012sacred}.

Taken jointly, the proposed model provides a reproducible, hypothesis-generating account of how role-structured conversion, retention, and schism interact in a shared population. Its contribution is a transparent dynamical framework, verified implementation, and a clear separation between endogenous mechanisms, prescribed forcing, and descriptive historical agreement. The results do not establish a universally predictive model of religious change or identify political causes. They instead specify testable mechanisms and the additional measurements needed to evaluate those mechanisms empirically.

\section*{Declarations}
\subsection*{Funding}
This study received no funding.
\subsection*{Data availability}
The numerical input series, source definitions, access dates, and provenance notes used in this revision accompany the code. Public sources are cited in the text. All historical observations are aggregate data; no individual-level records are used.
\subsection*{Code availability}
The entire code of the project can be freely found in \url{https://github.com/bilgesi/religion-dynamics-ode-abs}
\subsection*{Conflicts of interest}
The authors declare no conflict of interest.
\subsection*{Author contributions}
Bilge Taskin: Software, Formal analysis, Investigation, Data curation, Writing -- review and editing, Visualization. Teddy Lazebnik: Conceptualization, Methodology, Software, Validation, Writing -- original draft, Writing -- review and editing, Visualization, Supervision.

\bibliographystyle{unsrt}
\bibliography{biblio}

\clearpage

\FloatBarrier
\appendix
\section{Computation, diagnostics, and reproducibility}
\label{app:computation}
\subsection{Exchangeable-agent stochastic computation}
The fixed state space contains $1+3m$ compartments. Mutation does not make the dimension dynamic and does not prevent ordinary numerical integration. Stochastic simulation is used instead to quantify finite-population variability and discrete seeding, rather than to remedy an ill-defined ODE. This distinction is consistent with comparing equation-based and individual-based computational descriptions \cite{macal2010agent,schimit2023multi,scheidegger2018comparison,hunter2020hybrid,lazebnik2025comparing}.

At the start of each step, let $Q_{ij}(x,t)$ be the per-person hazard of transition from state $i$ to a distinct state $j$, and let an additional destination represent death. Its entries are exactly the conversion, role-change, disaffiliation, mutation, and death rates of Eqs.~\eqref{eq:S-dynamics}--\eqref{eq:P-dynamics}. For $H_i=\sum_jQ_{ij}$, the number leaving compartment $i$ is sampled as
\begin{equation}
L_i\sim\operatorname{Binomial}\{X_i,1-\exp(-H_i\Delta t)\}.
\end{equation}
Conditional destinations follow a multinomial distribution with weights $Q_{ij}/H_i$. This enforces at most one competing transition or death per individual per step, rather than independently applying mutually incompatible role changes. Independent births entering $i$ are $\operatorname{Poisson}(bX_i\Delta t)$, calculated using start-of-step counts. All changes are applied simultaneously after hazards have been computed. Under homogeneous mixing, this count-based sampler is equivalent to the corresponding exchangeable individual-agent update; it does not introduce a contact network.

The method exactly samples the specified frozen-hazard discrete-time process, but is not an exact continuous-time Markov-chain algorithm. Births and sequential within-step events introduce finite-step approximation relative to the continuous model. Scheduled introductions transfer existing people from $S$ and are applied exactly once at their stated grid times. The ODE is integrated to each actual event endpoint before the transfer or rate change; saved event-time states are right-continuous. Both computations use the same rates, event list, and compartment ordering.

For each replicate we calculate
\begin{equation}
y_r(t)=\frac{B_r(t)+M_r(t)+P_r(t)}{N(t)},
\label{eq:total-share}
\end{equation}
and only then average over replicates. This is not replaced by a ratio of mean counts. The displayed bands are pointwise empirical replicate quantiles. The master random seed is 20261022, scenario seeds are separated by 1000, and per-replicate seeds are derived using NumPy \texttt{SeedSequence}; exact seed arrays are saved with the trajectories. No run is selected or omitted on the basis of its outcome.

\subsection{ODE--stochastic comparison}
The comparison contains exactly 30 configurations: ten perturbed versions of each of A, B, and C, with $T=216$ years and $\Delta t=0.1$ throughout. Each nonzero behavioral rate and each $q_r$ is multiplied by an independent draw from $\operatorname{Uniform}(0.7,1.3)$; probabilities are clipped to $[0,1]$. Nonzero mutation entries are independently perturbed by the same rule, so A and C retain mutation and B retains zero mutation. Initial conditions, demographic rates, and event times are fixed. B's pre- and post-change values of a given religion-specific rate share the same perturbation multiplier. Each configuration uses 20 independent stochastic replicates.

MAE, RMSE, and $R^2=1-\sum_t(y^{\rm ABS}_t-y^{\rm ODE}_t)^2/\sum_t(y^{\rm ODE}_t-\bar y^{\rm ODE})^2$ are calculated per religion on a common grid, using the deterministic series as reference. Case-level values are unweighted means across religions, and reported summary means and standard deviations are across the 30 cases. They are not confidence intervals for a single case. All denominators in these reported $R^2$ computations are nonzero. The 30 configurations, per-religion metrics, and case-level metrics accompany the figure.

Table~\ref{tab:convergence} compares steps $0.1$, $0.05$, and $0.025$ years on a common annual output grid. Both stochastic sampling and finite-population differences contribute to the discrepancy from the deterministic solution. Accordingly, independent Monte Carlo errors do not decrease monotonically with step size; these checks do not establish equality of a finite-population mean and the nonlinear ODE. Conservation, boundary behavior, absent-religion invariance, event-state continuity, and agreement with an independently written RHS are additionally covered by executable regression tests.
\begin{table}[!htbp]\centering\small
\caption{Finite-step stochastic checks on a common annual grid over 216 years, using 100 replicates per row. MAE is against the ODE; MCSE is the mean pointwise standard error of the replicate mean. Both are fractions. These independently sampled errors need not decrease monotonically with step size.}\label{tab:convergence}
\begin{tabular}{lccc}\toprule Scenario & $\Delta t$ (years) & MAE & Mean pointwise MCSE\\\midrule
A & 0.100 & 0.001273 & 0.001076\\
A & 0.050 & 0.002118 & 0.000925\\
A & 0.025 & 0.002269 & 0.001094\\
B & 0.100 & 0.000217 & 0.000423\\
B & 0.050 & 0.000191 & 0.000430\\
B & 0.025 & 0.000264 & 0.000414\\
C & 0.100 & 0.000624 & 0.001337\\
C & 0.050 & 0.000735 & 0.001325\\
C & 0.025 & 0.001438 & 0.001360\\
\bottomrule\end{tabular}\end{table}

\subsection{Coexistence and horizon diagnostics}
For Scenario C, solving the frequency equations gives the equilibrium
\begin{align}
(s,b_1,m_1,p_1,b_2,m_2,p_2)
&\simeq(0.084976,0.116574,0.055359,0.285579,\nonumber\\
&\hspace{24mm}0.116574,0.055359,0.285579).
\end{align}
The constrained Jacobian eliminates $s=1-\sum_r(b_r+m_r+p_r)$; its eigenvalues are approximately
$-0.101992$, $-0.080837$, $-0.027345$, $-0.016912$, $-0.002723$, and $-0.001000$ year$^{-1}$. The population-conservation zero direction is not mistakenly included in the stability test. This numerical local analysis supports the specific coexistence example, not global attraction or robustness of all symmetric settings. The full equilibrium residual, eigenvalues, ablations, and both sensitivity-map horizons are saved without rounding in the code outputs.

\section{Historical calibration details}
\label{app:fitting}
\subsection{Observation construction and source audit}
\begin{table}[!htbp]\centering\small
\caption{NZ input counts and the constructed all-resident share. Published percentages are retained separately, not substituted for the count ratio. The denominator includes people without a classified affiliation response.}\label{tab:nzdata}
\begin{tabular}{ccccc}\toprule Year & No-religion count & All-resident count & Constructed share (\%) & Published (\%)\\\midrule
2001 & 1,028,049 & 3,737,280 & 27.508 & 29.6\\
2006 & 1,297,104 & 4,027,947 & 32.203 & 34.6\\
2013 & 1,635,345 & 4,242,048 & 38.551 & 41.9\\
2018 & 2,264,601 & 4,699,755 & 48.186 & 48.2\\
2023 & 2,576,049 & 4,993,923 & 51.584 & 51.6\\
\bottomrule\end{tabular}\end{table}

NZ no-religion counts through 2018 come from the 2019 Stats NZ release; the 2023 count is from the 2024 census release. The 2001 all-resident total is from the government Te Ara historical census table. The 2006 total is the 2013 census count less the officially reported 2006--2013 increase; later totals are reported directly. Random rounding means some releases differ by three persons. We retain the documented release-specific counts rather than silently swapping later rounded versions. The published percentages in Table~\ref{tab:nzdata} are a separate sensitivity series because their response denominators and processing differ. They are not alternative estimates obtained merely by rounding the primary ratios. In particular, dividing all five numerators by the 2013 population, or summing overlapping affiliations as a denominator, would not define the series used here.

Swedish observations are the 52 percentages in the cited official membership table, with 1981 omitted rather than interpolated. Turkey observations consistently use the \emph{peak} column; the data package also retains the source population and numerator, making the small share values auditable. Peak activity is an administrative maximum over a service year, not a census of all adherents. The source-year references and the extent of primary-source checking are recorded in \texttt{data/provenance.json}.

\subsection{Exact solution and identifiable fitting}
For a segment satisfying $\dot y=a y-cy^2$, the state propagated from its initial value $y_0$ after duration $u$ is
\begin{equation}
y(u)=\frac{y_0e^{au}}{1+(cy_0/a)(e^{au}-1)},\quad a\ne0;
\qquad y(u)=\frac{y_0}{1+cy_0u},\quad a=0.
\end{equation}
For Eq.~\eqref{eq:histfit_reduced_ode}, $a=\beta-\rho$ and $c=\beta$; for Turkey's logistic curve, $a=r$ and $c=r/K$. Sweden's second segment starts at the first segment's predicted value at 2000. No observed breakpoint value enters as a second initial condition. Linear and exponential alternatives are $y_0+a(t-t_0)$ and $y_0e^{g(t-t_0)}$, respectively. Their fitted values remain within the share range over the evaluated periods; they are descriptive baselines, not bounded population models for unrestricted extrapolation.

Least squares uses residuals after the fixed initial observation and scales them by the training-series range only to improve numerical conditioning. Bounds are $0\leq\beta,\rho\leq1$ year$^{-1}$ in each relevant segment, $10^{-8}\leq r\leq1$ year$^{-1}$, and $1.000001\max(y_{\rm train})\leq K\leq1$. Logistic optimization uses $\log K$ internally but reports $K$. The exponential rate is bounded to $[-1,1]$ year$^{-1}$, and the anchored linear slope has its closed-form least-squares solution. Nonlinear optimization uses SciPy \texttt{least\_squares}, tolerances $10^{-12}$, and up to 2500 function evaluations per start. Starting values, bounds, successful-fit flags, and residual-Jacobian singular values are saved. Multiple starts are a robustness check, not a mathematical guarantee of a global optimum.

The Swedish recruitment estimates and the primary NZ loss estimate are at their nonnegative lower bounds. They must not be interpreted as measured zero conversion or zero return to religion. Aggregate shares confound omitted mechanisms, and small residual errors do not identify the full role structure. For the descriptive AICc diagnostic we use
\begin{equation}
\mathrm{AICc}=n\log(\mathrm{SSE}/n)+2k^*+
\frac{2k^*(k^*+1)}{n-k^*-1},\qquad k^*=k+1,
\end{equation}
where $k$ is the number of curve parameters, the extra parameter is an independent Gaussian error variance, and $n$ excludes initialization. Constants common to models fitted to the same observations are omitted. This working error model is not a substantive model of census uncertainty or temporal dependence. For the two-parameter NZ curve, $n=4$ makes the correction undefined; no finite AICc is reported. Table~\ref{tab:aicc} also supplies the fitted trend coefficients.
\begin{table}[!htbp]\centering\small
\caption{Descriptive complexity diagnostics and anchored trend coefficients. AICc assumes independent Gaussian errors, counts the error variance as a parameter, and is not reported when the small-sample correction is undefined. Comparisons across different datasets are not meaningful. $a$ is the anchored linear slope (fraction/year), and $g$ is the exponential rate (year$^{-1}$).}\label{tab:aicc}
\begin{tabular}{llrl}\toprule Case & Specification & AICc & Trend coefficient\\\midrule
Sweden & Constant rates & -280.992 & --\\
Sweden & Rate shift & -468.808 & --\\
Sweden & Linear trend & -298.236 & $a=-0.00676724$\\
Turkey & Logistic & -241.83 & --\\
Turkey & Exponential & -242.871 & $g=0.076951$\\
Turkey & Linear trend & -245.107 & $a=3.50551e-06$\\
NZ & Constant rates & -- & --\\
NZ & Linear trend & -17.498 & $a=0.0110103$\\
\bottomrule\end{tabular}\end{table}

\begin{figure}[!htbp]\centering
\begin{subfigure}[t]{.48\textwidth}\includegraphics[width=\linewidth]{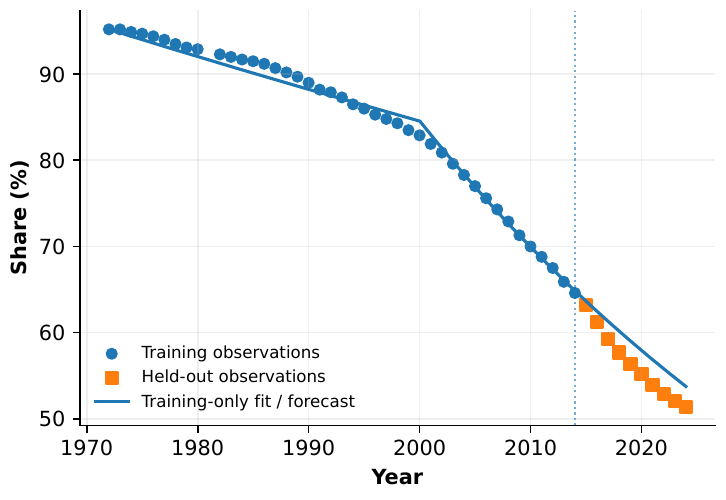}\caption{Sweden; train through 2014.}\end{subfigure}\hfill
\begin{subfigure}[t]{.48\textwidth}\includegraphics[width=\linewidth]{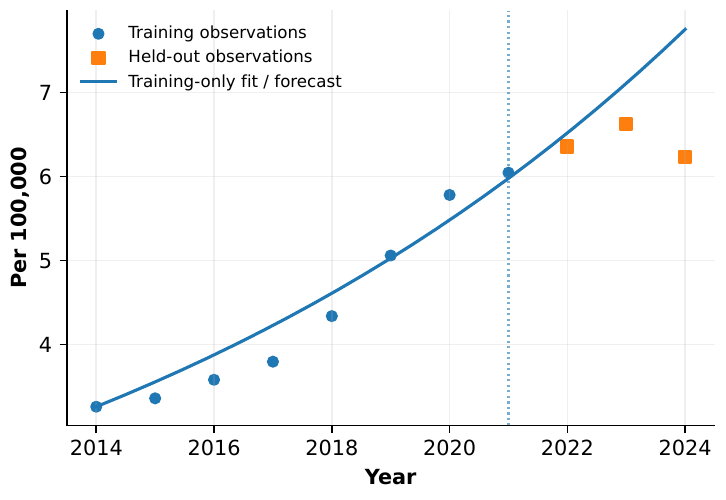}\caption{Turkey; train through 2021.}\end{subfigure}\par\medskip
\begin{subfigure}[t]{.48\textwidth}\includegraphics[width=\linewidth]{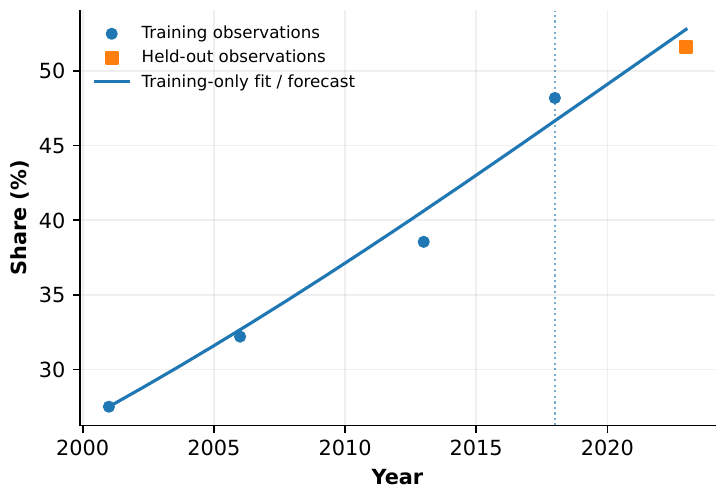}\caption{NZ; train through 2018.}\end{subfigure}
\caption{Genuine held-out evaluations of the selected reduced dynamical specifications. The curves use training-only parameter estimates on both sides of the vertical cutoff. Square points were not used for fitting, breakpoint selection, or state reinitialization. Full-data fits in Figure~\ref{fig:historic_fit} are not substituted for these forecasts.}\label{fig:holdout}
\end{figure}

\end{document}